\documentclass[journal,twocolumn]{IEEEtran}
\usepackage{cite}
\ifCLASSINFOpdf
   \usepackage[pdftex]{graphicx}
\else
\fi

\usepackage{multirow}

\begin{document}
%
% paper title
% Titles are generally capitalized except for words such as a, an, and, as,
% at, but, by, for, in, nor, of, on, or, the, to and up, which are usually
% not capitalized unless they are the first or last word of the title.
% Linebreaks \\ can be used within to get better formatting as desired.
% Do not put math or special symbols in the title.
\title{ Seed Layer Engineering for Crack-free Sol-gel Alumina Deposition on GFETs }
%
%
% author names and IEEE memberships
% note positions of commas and nonbreaking spaces ( ~ ) LaTeX will not break
% a structure at a ~ so this keeps an author's name from being broken across
% two lines.
% use \thanks{} to gain access to the first footnote area
% a separate \thanks must be used for each paragraph as LaTeX2e's \thanks
% was not built to handle multiple paragraphs
%

\author{Nama Premsai, 
Department of Electrical engineering, Indian Institute of Technology Bombay, Mumbai-76, e-mail: premsainama@ee.iitb.ac.in, premsaii45@gmail.com.
~\IEEEmembership{}
~\IEEEmembership{}
        % <-this % stops a space
%\thanks{Nama Premsai is with the Department
%of Electrical Engineering, Indian Institute of Technology, Mumbai,
%400076, India,  e-mail: (see premsainama@ee.iitb.ac.in).}% <-this % stops a space
%\thanks{J. Doe and J. Doe are with Anonymous University.}% <-this % stops a space
%\thanks{Manuscript received April 19, 2005; revised August 26, 2015.}

}

\maketitle 

% As a general rule, do not put math, special symbols or citations
% in the abstract or keywords.
\begin{abstract}

Low cost and low thermal budget based spin-coated sol-gel Alumina was explored as a dielectric/passivation layer for GFET. Post thermal annealing, the crack was observed in sol-gel Alumina layer exactly above the graphene channel. The possible mechanism of crack could be graphene lateral restoring movement due to (i) Thermal Expansion Coefficient (TEC) difference between graphene and adjacent layers and (ii) shrinkage stress generated during the solvent removal process. Based on the crack formation phenomenon, a combination of different annealing schemes (low thermal budget DUV annealing) and seed layer engineering (thickness and different deposition schemes) were carried out. Finally, a novel two-step seed layer deposition method with DUV annealing was proposed and demonstrated to resolve the crack issue successfully and also able to retain the Dirac point in the electrical characteristics. 
\end{abstract}

% Note that keywords are not normally used for peerreview papers.

% For peer review papers, you can put extra information on the cover
% page as needed:
% \ifCLASSOPTIONpeerreview
% \begin{center} \bfseries EDICS Category: 3-BBND \end{center}
% \fi
%
% For peerreview papers, this IEEEtran command inserts a page break and
% creates the second title. It will be ignored for other modes.
\IEEEpeerreviewmaketitle

\section{Introduction}

 \IEEEPARstart{T}{he} uniqueness in the properties of graphene like am-bipolar transport, high carrier mobility \cite{novoselov2004electric}, exceptional mechanical strength \cite{lee2012multi}, thermal properties \cite{balandin2011thermal} and flexible nature \cite{park2016extremely} have paved the way for the RF applications over conventional materials \cite{han2014graphene}.
Dielectric materials plays an important role of gate dielectric and passivation layer for any FET devices. There are different techniques like PVD \cite{wu2008top,kedzierski2008epitaxial} and ALD \cite{wang2008atomic,kim2009realization,fallahazad2012scaling,kang2013mechanism} to deposit high quality dielectric layers. On the other hand there are reports of solution based deposition techniques through spin or spray coatings for dielectric materials \cite{park2017sol}. These techniques have advantages of high deposition rate, no vacuum requirement, low cost method, tuning of the precursor composition \cite{brinker2013sol} and can be used for flexible applications \cite{wang2018high}. But there are also several challenges in using solution based deposition techniques like (i) adhesion of spray or spin coated layer with the underneath layer and (ii) need to anneal layers post deposition to remove solvent from the layer, which leads to stress in the films. 

The most commonly used dielectrics for the GFETs are Si\textsubscript{3}N\textsubscript{4} \cite{zhu2010silicon}, HfO\textsubscript{2} \cite{xiao2017atomic} and Al\textsubscript{2}O\textsubscript{3} \cite{schiliro2019recent}. Among them, the Al\textsubscript{2}O\textsubscript{3} has proven to be the best passivation layer due to its high thermal stability, appreciable permeability barrier and reduced hysteresis in transfer characteristics in FET devices \cite{zurutuzaaelorza2015highly}. The presence of the Al\textsubscript{2}O\textsubscript{3} monolayer does not introduce interface states and electron-hole puddles in graphene \cite{song2016stability}. Recently the sol-gel Alumina (Al\textsubscript{2}O\textsubscript{3}) has been extensively used in the solar cells as a field-effect passivation layer \cite{srinivasan2019aluminium} due to its good interface quality and dielectric breakdown. Park et.al\cite{park2016solution,kim2019direct} and Bae et.al\cite{bae2014fabrication} have explored sol-gel Alumina on the graphene transistors as dielectric layer and pH sensing membrane.

Before sol-gel Alumina spin coating, the surface of graphene needs to be made hydrophilic for its proper adhesion. There are few techniques like plasma exposure \cite{shin2010surface}, functionalisation \cite{singla2019stable}, seed layer deposition \cite{park2016solution,wang2008microwave}  explored for wetting the graphene surface. Techniques such as O\textsubscript{2} plasma exposure, UV-O\textsubscript{3} irradiation and functionalization would create lattice damage \cite{imamura2015modification} and doping to the graphene\cite{singla2019stable}. On the other hand, the seed layer deposition does not create any lattice damage and also won’t create any P-type doping \cite{fallahazad2012scaling, feng2012unipolar, shi2014selective}. The deposition of the seed layer is usually carried out by the e-beam evaporation technique as it provides structural and morphological control of films and offers low resistivity \cite{harsha2005principles}. The e-beam lacks uniform deposition in the inner surface of 3D complex geometries due to the shadowing effect caused by line of sight deposition \cite{volmer2021solve}. Sputtering can also be used as an alternative for deposition of a seed layer for conformal coverage of steps and trenches \cite{brown2015impact}. The problem associated with sputtering was that it can damage the underlying surface with plasma.

The proper densification of sol-gel thin film requires an annealing process to remove the solvent and also some trapped byproducts from the oxide network \cite{bae2014fabrication, park2016solution}. The conventional method i.e thermal annealing mostly requires elevated temperature which can cause strain \cite{lee2012optical} and unintentional doping \cite{sojoudi2012impact} in graphene devices. Also, during the annealing of the sol-gel solution, the drying stress due to the differential shrinkage (capillary forces) of the gel will be present \cite{brinker2013sol,HiromiKozuka2003Sol}. But at high temperatures, the films are prone to undergo in-plane stress i.e tensile stress during the heating and compressive stress after cooling \cite{kozuka2003stress,kozuka2006stress}. To reduce annealing temperature, techniques such as Deep Ultra Violet (DUV) annealing \cite{kim2012flexible, park2015depth, jo2017ultralow}, O\textsubscript{3} annealing \cite{huang2021high}, Microwave annealing \cite{wang2008microwave,kim2017solution,kim2019direct,kim2014tuning}, High pressure annealing \cite{rim2012simultaneous} are mostly used. Among these DUV annealing process was the most explored and readily available technique for sol-gel Alumina \cite{park2015depth}.

In this work, we report fabrication of sol-gel alumina based GFET devices. We discovered crack formation in the sol-gel Alumina layer post annealing. We explored seed layer engineering and different annealing techniques (Thermal and DUV) based on understanding different phenomenons to resolve the crack issue. Finally we propose novel two step method to resolve the crack issue.

\section{Experimental}
\subsection{Synthesis of sol-gel Alumina}
The synthesis of sol-gel Alumina was carried out by mixing the Aluminum nitrate nonahydrate (Al(NO\textsubscript{3})\textsubscript{3}.9H\textsubscript{2}O) (from Sigma Aldrich) with CMOS grade Isopropyl alcohol (C\textsubscript{3}H\textsubscript{7}OH). The mixture was stirred for 12 hours and the resultant solution was filtered using a membrane filter with a pore size of 0.22 $\mu$m. The molarity of the resultant solution was kept at 0.1 M. The resultant solution was spin-coated on samples at 5000 rpm for 45 seconds. Thermal \cite{park2016solution} and DUV annealing \cite{park2015depth} techniques have been used to remove solvent from spin-coated sol-gel Alumina.

\subsection{Fabrication of GFETs}
\begin{figure}[!ht]
\centering
\includegraphics[width= \columnwidth]{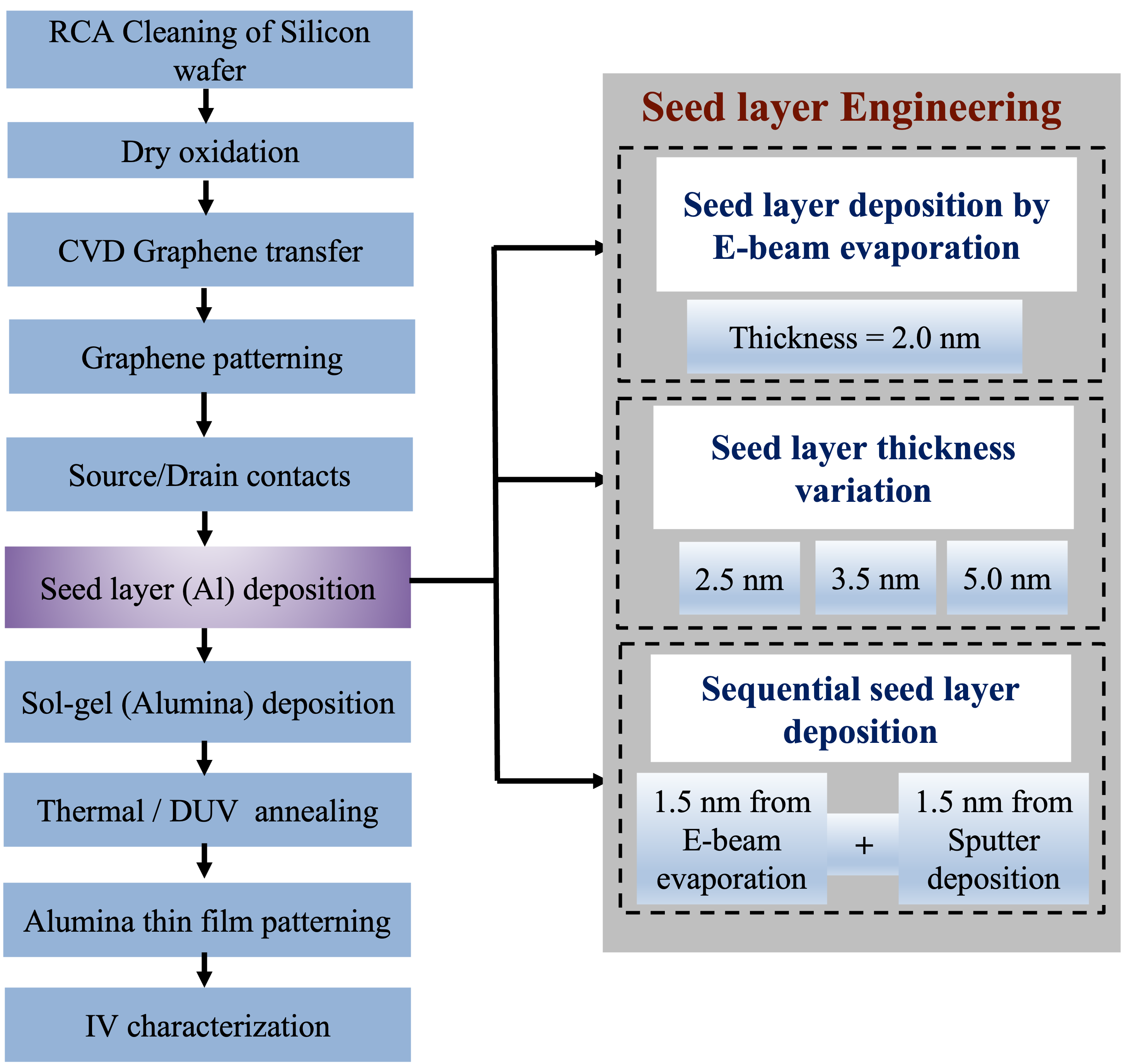}
\caption{\label{expt-split} The complete fabrication flow for GFETs with different splits for seed layer engineering}
\end{figure}

The fabrication flow of graphene-based transistor with a seed layer and passivation layer is shown in Figure \ref{expt-split}. To fabricate the graphene transistors, CVD-grown graphene monolayer on copper foil were purchased from Graphenea Inc. These copper foil was cut into pieces and coated with 2$\%$ PMMA 950K. Next, the pieces were immersed in the Copper etchant solution. The isolated graphene layers were carefully transferred to a thermally oxidized 2-inch silicon substrate with a resistivity of 0.002-0.005 ohm-cm. The thickness of the SiO\textsubscript{2} was kept at 90 nm to make the transferred graphene layer optically visible \cite{blake2007making}. The backside oxide was etched using 7:1 buffered Hydrofluoric (BHF) acid solution for the substrate contact. Then, a graphene monolayer was patterned for the graphene channels using the optical lithography followed by the oxygen plasma exposure. The device dimensions used are width (W = 5-7$\mu$m) and length (L = 20-30$\mu$m). After channel definition, the source and drain contacts were made by using the electron beam evaporation (e-beam) of 10 nm Nickel and 40 nm Gold followed by the lift-off process. Thereafter, the Aluminum seed layer engineering was carried out as shown in Figure \ref{expt-split}. Subsequently, the deposited seed layers were subjected to natural oxidation in the ambient atmosphere for one day. Next, the synthesized sol-gel Alumina was spin-coated followed by the thermal/DUV annealing. Finally, the densified sol-gel Alumina film was patterned to complete the device fabrication. The electrical characterization on the respective samples ware carried out done using a Keysight B1500A semiconductor device analyzer.

\section{Results and discussion}

\subsection{Thin seed layer deposition by e-beam evaporation }
\begin{figure}[!ht]
\centering
\includegraphics[width= \columnwidth]{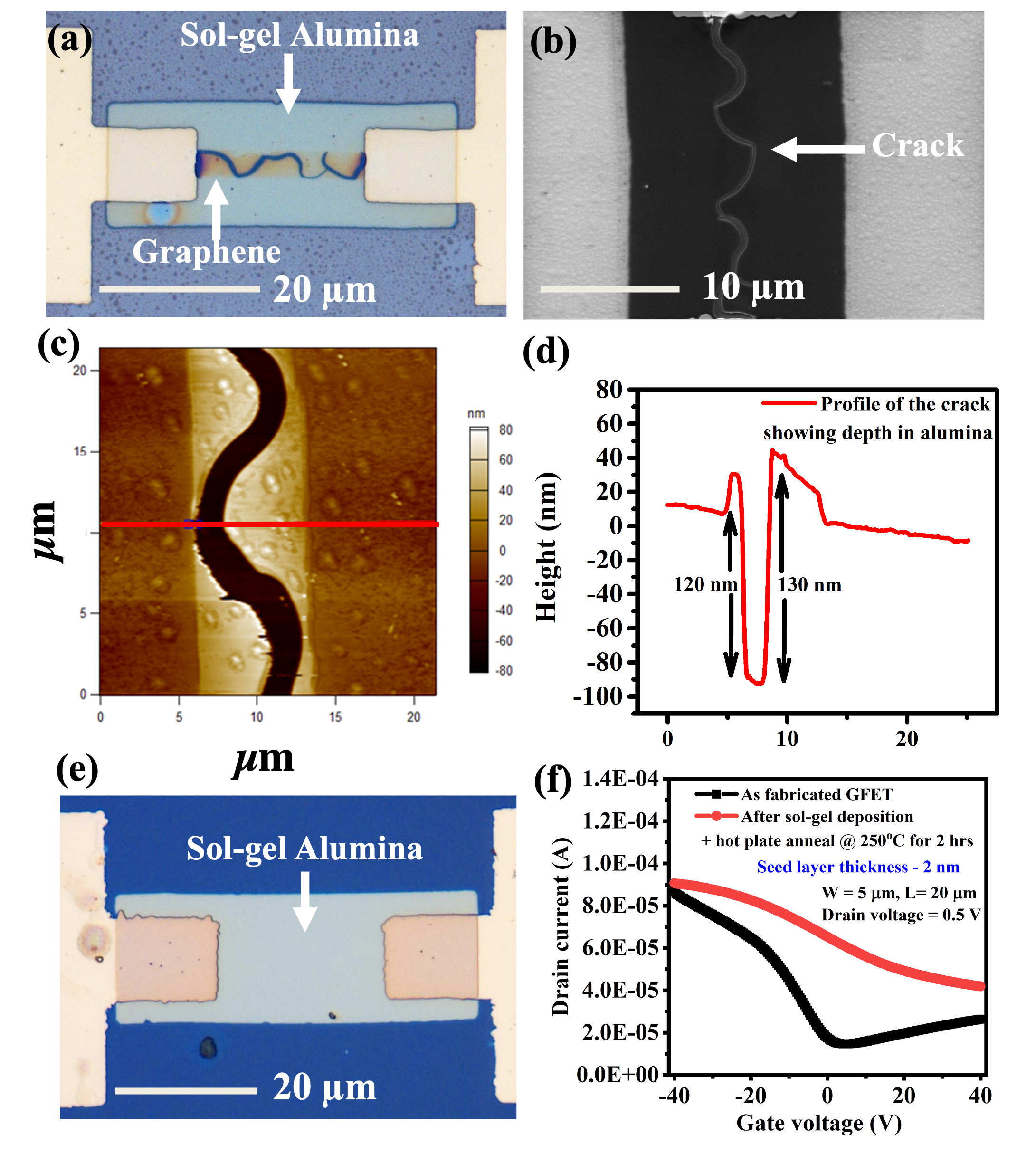}
\caption{\label{first-deposition}(a) Microscope image of GFET after sol-gel densification at 250\textsuperscript{o}C for 2 hours (b) SEM image of the GFET showing the crack on the graphene channel in sol-gel Alumina. (c) AFM image of the crack (d) Depth profiling of the crack (e) Microscope image of sol-gel Alumina without graphene on SiO\textsubscript{2} (f) Transfer characteristics of the GFET of Figure (a)}
\end{figure}

Figure \ref{first-deposition} shows the effect of thermal annealing on GFET with a seed layer thickness of 2 nm deposited using e-beam evaporation. The annealing was carried out on a hotplate at 250\textsuperscript{o}C for 2 hours. During the densification by thermal annealing, the crack was observed in the sol-gel Alumina layer of GFET exactly above the graphene channel layer as shown in Figure \ref{first-deposition} (a) and (b). The Atomic Force Microscopy (AFM) image (Figure \ref{first-deposition} (c)) with profiling  (Figure \ref{first-deposition} (d)) conforms that crack formation in the sol-gel Alumina layer. The depth profile shows a non-uniformity at the edges of the cracks due to the agglomeration of the Alumina film after densification.

The origin of crack formation was investigated by fabricating the entire device without a graphene layer. It was found that there was no crack formation without a graphene layer (Figure \ref{first-deposition}(e)). This confirms that the graphene layer plays a major role in crack formation in the sol-gel Alumina layer during the annealing process. There could be the following two phenomenons happening during annealing, (i) solvent removal and (ii) expansion and/or compression of material depending upon the TEC of the material. There are reports of stress generated due to shrinkage of sol-gel Alumina during annealing \cite{gawel2010sol,brinker2013sol}. Positive TEC leads to expansion of material and negative annealing leads to compression during annealing. Graphene has a negative TEC \cite{yoon2011negative}, on other hand SiO\textsubscript{2} substrate and oxidized seed layer has positive TEC \cite{bao2009controlled,wilson1941thermal}. There are reports that when there was a negative TEC graphene on positive TEC SiO\textsubscript{2} substrate, then it leads to slippage (lateral restoring movement) of graphene during annealing due to TEC difference between them \cite{yoon2011negative}. The threshold temperature to cause slippage was reported to be 117\textsuperscript{o}C \cite{jiang2018equi}. Both of the above phenomena might lead to the movement of graphene, as graphene was residing on the substrate with  van der Waals forces. The movement of graphene might lead to crack formation in the sol-gel Alumina film.

The Figure \ref{first-deposition} (f) shows the transfer characteristics of GFET for device Figure \ref{first-deposition}(a). The output drain current after the thermal annealing of sol-gel Alumina shows the disappearance of the Dirac point and the electron branch. This conforms to the P-type nature of the graphene \cite{guo2011graphene}. This is due to the diffusion of O\textsubscript{2} and H\textsubscript{2}O from the ambient atmosphere which will form the redox reaction with the graphene during annealing \cite{kang2013mechanism} in addition to the stress. These redox reactions will form the OH\textsuperscript{-}ions at the interface which will act as coulomb scatters. 

\begin{figure}[!ht]
\centering
\includegraphics[width= \columnwidth]{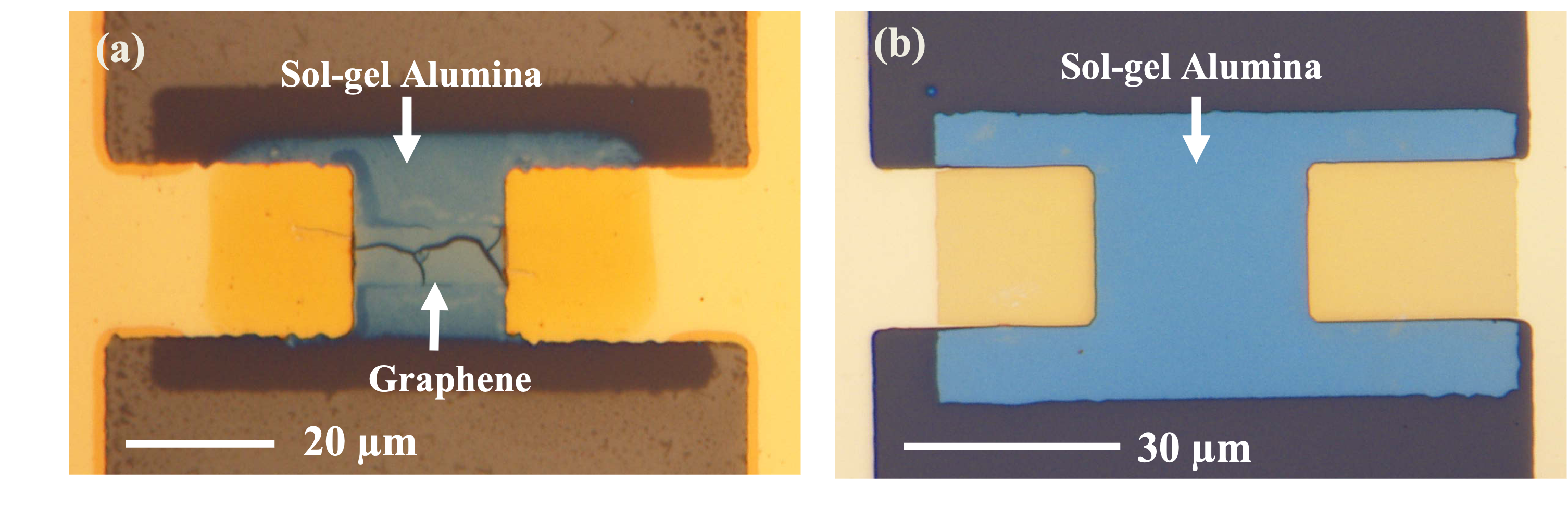}
\caption{\label{duv-first}(a) Microscope image of GFET after sol-gel densification by DUV annealing for 30 mins with graphene channel (b) Without the graphene channel on SiO\textsubscript{2}}
\end{figure}

The TEC component can be reduced by using an annealing technique with a lower thermal budget. There are reports of DUV annealing to reduce thermal budget during annealing \cite{park2015depth}. The DUV annealing technique has been applied in this work. The temperature was measured during DUV annealing using thermo-couple and it was found to be close to 70\textsuperscript{o}C. The graphene device fabricated using DUV annealing still exhibits crack formation in the sol-gel layer exactly above the graphene layer as shown in the Figure \ref{duv-first}(a). On the other hand, the device without graphene exhibits no crack as shown in the Figure \ref{duv-first}(b). Hence it can be concluded that reducing the thermal budget was not sufficient to avoid crack formation. It means that even with DUV annealing the movement of graphene was not restricted. 

\subsection{Seed layer thickness variation}

\begin{figure}[!ht]
\centering
\includegraphics[width= \columnwidth]{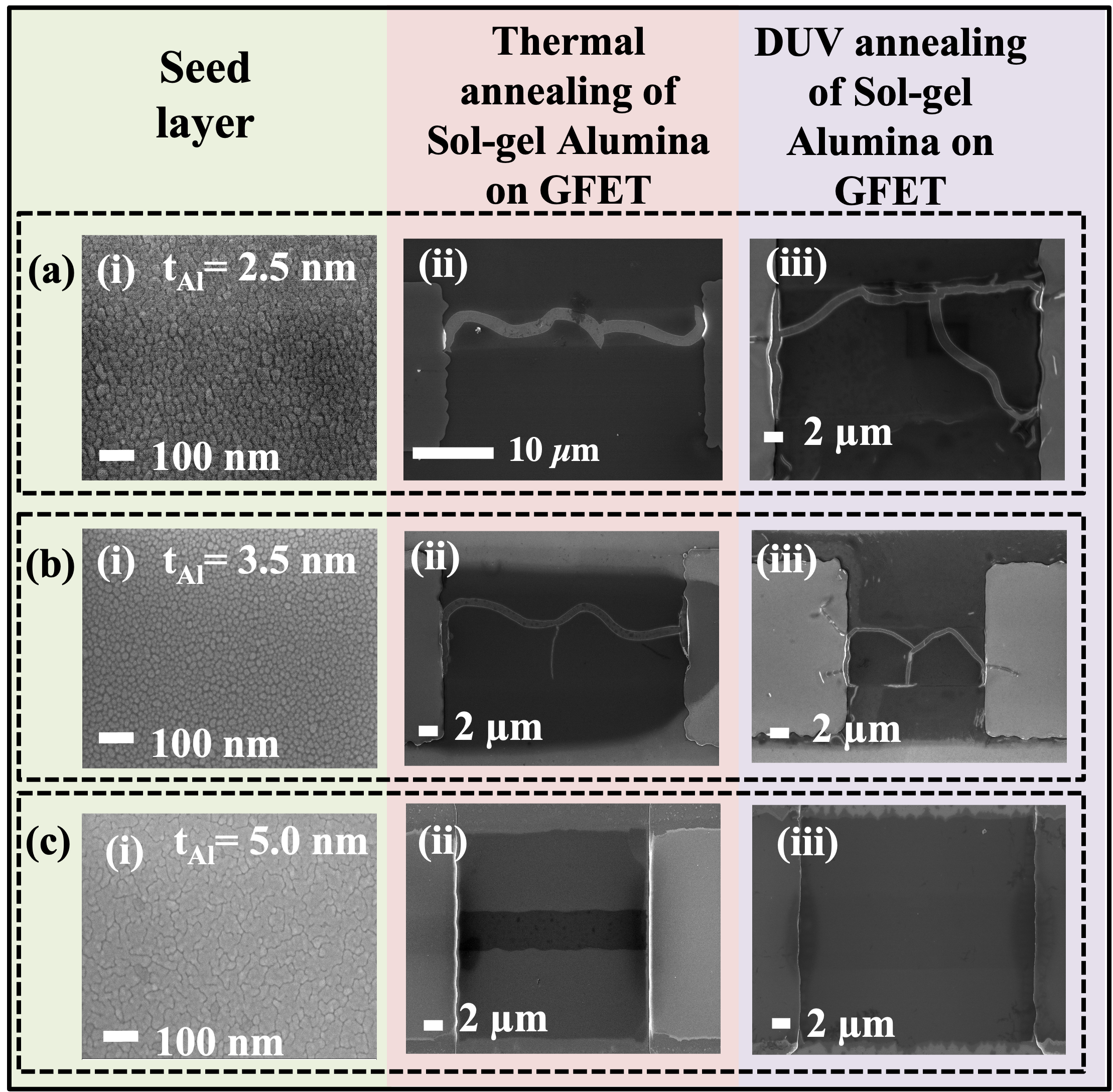}
\caption{\label{seedlayervar} The study of the crack in Alumina thin film on graphene channel by seed layer thickness variation on GFET before sol-gel deposition. (a), (b) and (c) are three sets used to study the seed layer thickness variation represented by dashed lines. Subfigure (i) in (a), (b), and (c) represent the micrograph of the Aluminum seed layer of thickness 2.5 nm, 3.5 nm, and 5.0 nm.  Subfigure (ii) in (a),(b), and (c) represents SEM images after the thermal annealing of Alumina on GFET. The subfigure (iii) in (a), (b), and (c) represents SEM images after the DUV annealing of Alumina on GFET.}
\end{figure}

\begin{figure}[!ht]
\centering
\includegraphics[width=\columnwidth]{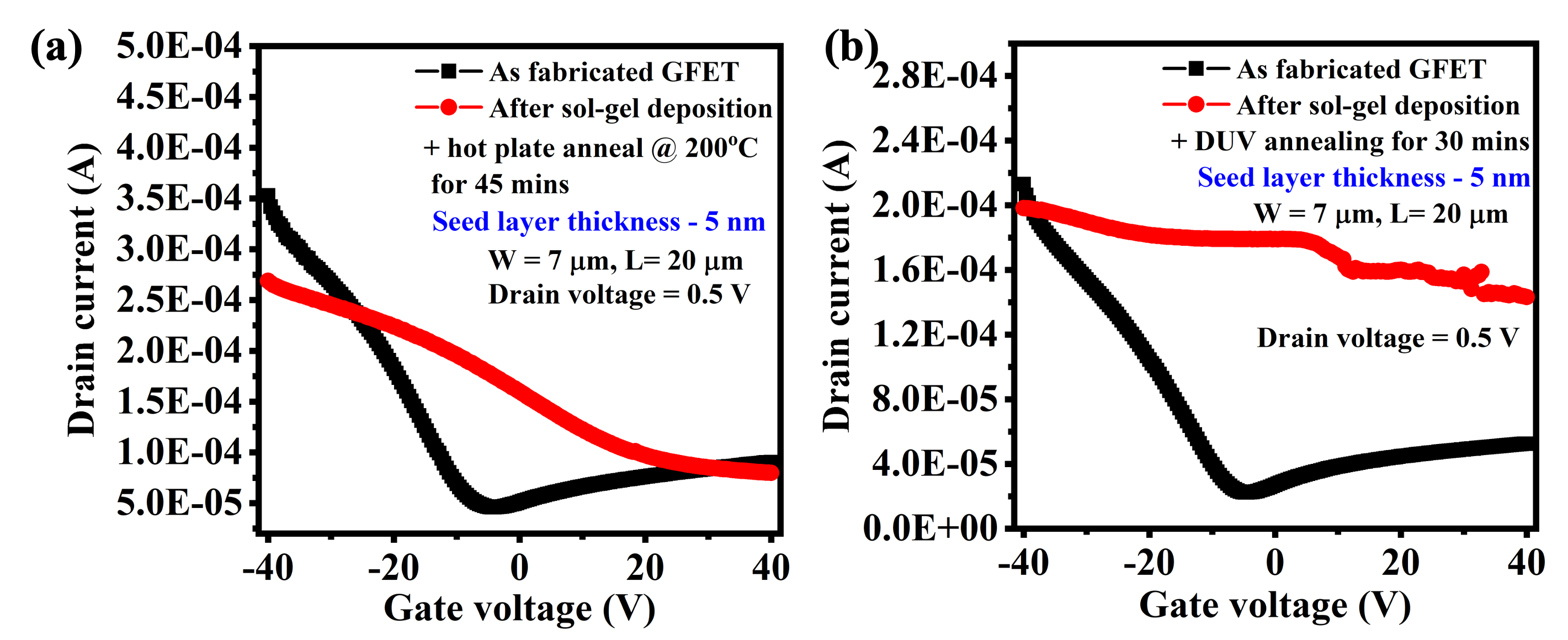}
\caption{\label{IV5nm} Transfer characteristics of graphene FETs after (a) thermal and (b) DUV annealing of sol-gel Alumina with 5 nm seed layer thickness }
\end{figure}

To further reduce graphene movement, different thicknesses (2.5, 3.5, and 5 nm) of seed layers were tried to make a graphene device. As shown in subfigures (i)-(iii) of Figure \ref{seedlayervar}(a) and (b), it can be seen that graphene devices with 2.5 and 3.5 nm exhibit crack formation. On the other hand, 5 nm device does not show any crack formation as shown in Figure \ref{seedlayervar}(c)(i)-(iii). Hence devices with thicker seed layers were able to restrict graphene movement to avoid crack formation. But when the graphene devices with a 5 nm seed layer were electrically tested before and after the sol-gel deposition as shown in the Figure \ref{IV5nm}(a) and (b). It was found that post-sol-gel deposition, devices show the disappearance of the Dirac point and resulted in the hole doping after the thermal annealing. In contrast, the drain current after DUV annealing did not show any modulation with gate voltage variation which means that the Aluminum seed layer could not have fully oxidized. Further deeper analysis of fabrication pointed out that a thin seed layer (~2.5 nm) was deposited using e-beam evaporation, which was the line of sight deposition and might not lead to side-wall coverage (Figure \ref{seedtype-2}(a)).

\subsection{Sequential seed layer deposition}

\begin{figure}[!ht]
\centering
\includegraphics[width= \columnwidth]{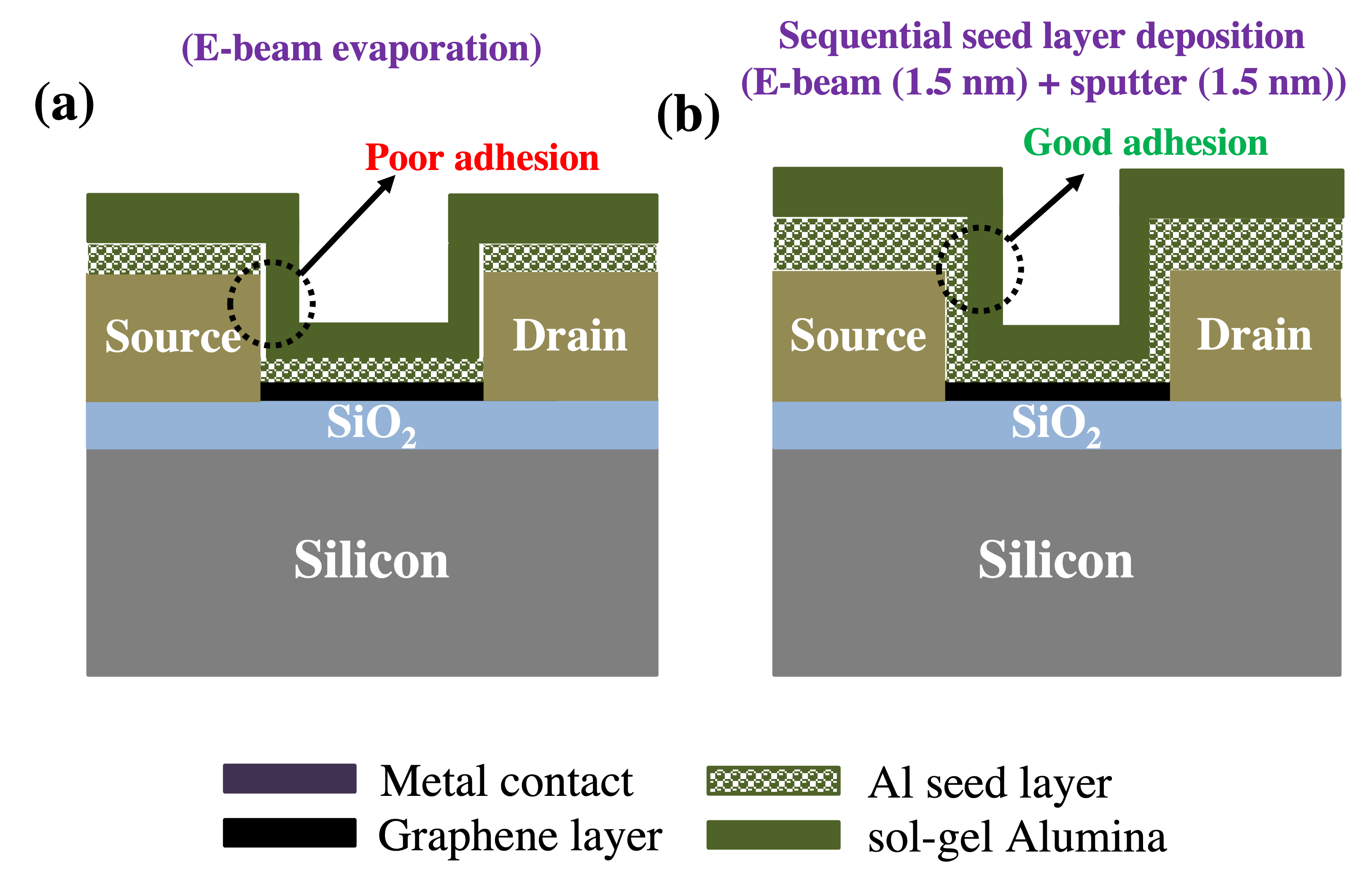}
\caption{\label{seedtype-2} 2-D Schematic of (a) GFET with seed layer type-1 deposited from e-beam evaporation (the lack of seed layer on side walls) (b) GFET with seed layer type-2 deposited in sequence from e-beam (1.5 nm) evaporation and sputter (1.5 nm) (for good side wall deposition)}
\end{figure}

To avoid insufficient oxidation and crack formation issues, we propose a novel two-step seed layer deposition approach. (i) first thin seed layer deposition using e-beam evaporation to avoid damage to graphene and (ii) thin seed layer deposition using sputter process to have sidewall coverage. The device was fabricated with a two-step process as shown in Figure \ref{seedtype-2}(b)

\begin{figure}[!ht]
\centering
\includegraphics[width= \columnwidth]{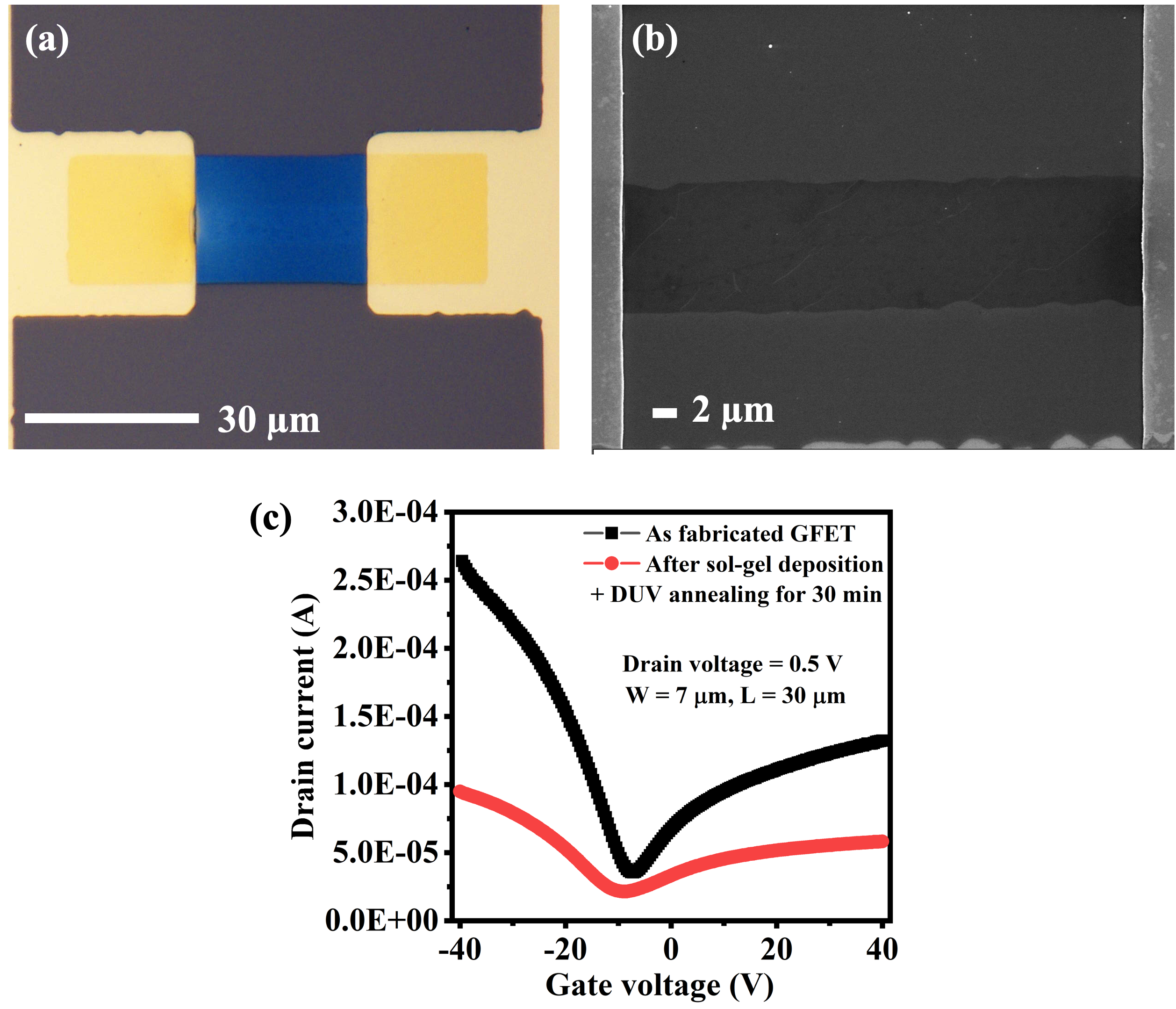}
\caption{\label{withoutcrack} (a) Microscope image of GFET after sol-gel Al\textsubscript{2}O\textsubscript{3} deposition and DUV annealing with seed layer type-2 prior to sol-gel deposition (b) SEM image of GFET after sol-gel Alumina deposition and DUV annealing with seed layer type-2 prior to sol-gel deposition (c) Transfer characteristics of GFET before and DUV annealing of sol-gel Alumina }
\end{figure}

As shown in the Figure \ref{withoutcrack} (a) and (b), no cracks were observed in Alumina film on the graphene channel region. The sequential deposition will provide dense, hydrophilic, assist in the faster heat transfer from the metal contacts to the sol-gel Alumina, and also withstands the stress-induced slippage generated by TEC mismatch and stress due to shrinkage of the sol-gel Alumina during annealing as shown in the schematic \ref{seedtype-2}(b)

Figure \ref{withoutcrack} (c) shows the transfer characteristics for the As fabricated GFET and after the DUV annealing with sequential seed layer deposition. The As fabricated GFET shows the Dirac point at around -6.4 V which was the signature of the n-type doping \cite{guo2011graphene}. This shift could be due to long exposure of the resist stripper during lift-off process \cite{sul2016reduction}. The asymmetry in transfer characteristics was due to the work-function difference between Nickel (Work function - 5.01 eV) and graphene (Work function - 4.5 eV) \cite{nagashio2009metal}. However, the GFET retains its Dirac point after the DUV annealing in contrast to thermal annealing but with a slight shift in Dirac point towards negative voltage which is at around -8 V. The mobility of the GFET before and after DUV annealing was calculated using the FTM method \cite{kim2009realization}. The extracted parameters and their comparison are reported in the table \ref{mobilityextr}

\begin{table}
\centering
\caption{Mobility, Residual carrier concentration, Contact resistance extracted using FTM method before and after DUV annealing}
\begin{tabular}{|c|cl|cl|}
\hline
\multirow{2}{*}{\textbf{Parameters}}                                                        & \multicolumn{2}{c|}{\textbf{\begin{tabular}[c]{@{}c@{}}As fabricated\\ GFET\end{tabular}}}                                                               & \multicolumn{2}{c|}{\textbf{\begin{tabular}[c]{@{}c@{}}After sol-gel \\ deposition\\ +\\ DUV \\ annealing\end{tabular}}}                                 \\ \cline{2-5} 
                                                                                           & \multicolumn{1}{c|}{\textbf{\begin{tabular}[c]{@{}c@{}}Hole\\ branch\end{tabular}}} & \textbf{\begin{tabular}[c]{@{}l@{}}Electron\\ branch\end{tabular}} & \multicolumn{1}{c|}{\textbf{\begin{tabular}[c]{@{}c@{}}Hole\\ branch\end{tabular}}} & \textbf{\begin{tabular}[c]{@{}l@{}}Electron\\ branch\end{tabular}} \\ \hline
\textbf{\begin{tabular}[c]{@{}c@{}}Mobility\\ (cm\textsuperscript{2}/V-sec)\end{tabular}}                    & \multicolumn{1}{c|}{4400}                                                           & 4281                                                               & \multicolumn{1}{c|}{1174}                                                           & 1626                                                               \\ \hline
\textbf{\begin{tabular}[c]{@{}c@{}}Residual carrier\\ concentration\\ (/cm\textsuperscript{2})\end{tabular}} & \multicolumn{1}{c|}{5.2e11}                                                         & 8.9e11                                                             & \multicolumn{1}{c|}{7.1e11}                                                         & 1.1e12                                                             \\ \hline
\textbf{\begin{tabular}[c]{@{}c@{}}Contact \\ resistance\\ (ohms)\end{tabular}}            & \multicolumn{1}{c|}{1134}                                                           & 3399                                                               & \multicolumn{1}{c|}{2084}                                                           & 7169                                                               \\ \hline
\end{tabular}

\label{mobilityextr}
\end{table}

The decrease of the mobility in electron and hole branches after DUV annealing could be attributed to the charged impurity scattering \cite{chen2008charged}. The summary of the entire work is given in table \ref{summary}.

\begin{table}[]
\centering
\caption{Summary table for crack and Dirac point for thermal and DUV annealing (*Yes = Exists, No = Absent, and '--' = Unattempted )}
\label{summary}
\begin{tabular}{|c|c|cc|cc|}
\hline
\multirow{2}{*}{\textbf{\begin{tabular}[c]{@{}c@{}}Seed \\ layer\\ process\end{tabular}}}           & \multirow{2}{*}{\textbf{\begin{tabular}[c]{@{}c@{}}Thickness \\ (nm)\end{tabular}}} & \multicolumn{2}{c|}{\textbf{Crack}}                                                                                                                    & \multicolumn{2}{c|}{\textbf{Dirac point}}                                                                                                                \\ \cline{3-6} 
                                                                                                    &                                                                                     & \multicolumn{1}{c|}{\textbf{\begin{tabular}[c]{@{}c@{}}Thermal\\ anneal\end{tabular}}} & \textbf{\begin{tabular}[c]{@{}c@{}}DUV\\ anneal\end{tabular}} & \multicolumn{1}{c|}{\textbf{\begin{tabular}[c]{@{}c@{}}Thermal \\ anneal\end{tabular}}} & \textbf{\begin{tabular}[c]{@{}c@{}}DUV \\ anneal\end{tabular}} \\ \hline
\textbf{e-beam}                                                                                     & 2.0 nm                                                                              & \multicolumn{1}{c|}{Yes}                                                               & --                                                            & \multicolumn{1}{c|}{--}                                                                 & --                                                             \\ \hline
\multirow{3}{*}{\textbf{\begin{tabular}[c]{@{}c@{}}Thickness\\ variation\\ (e-beam)\end{tabular}}}  & 2.5 nm                                                                              & \multicolumn{1}{c|}{Yes}                                                               & Yes                                                           & \multicolumn{1}{c|}{--}                                                                 & --                                                             \\ \cline{2-6} 
                                                                                                    & 3.5 nm                                                                              & \multicolumn{1}{c|}{Yes}                                                               & Yes                                                           & \multicolumn{1}{c|}{--}                                                                 & --                                                             \\ \cline{2-6} 
                                                                                                    & 5.0 nm                                                                              & \multicolumn{1}{c|}{No}                                                                & No                                                            & \multicolumn{1}{c|}{No}                                                                 & No                                                             \\ \hline
\textbf{\begin{tabular}[c]{@{}c@{}}Sequential \\ seed layer  \\ (e-beam + \\ Sputter)\end{tabular}} & \begin{tabular}[c]{@{}c@{}}1.5 nm\\ + \\ 1.5 nm\end{tabular}                        & \multicolumn{1}{c|}{--}                                                                & No                                                            & \multicolumn{1}{c|}{--}                                                                 & Yes                                                            \\ \hline
\end{tabular}
\end{table}

\section{Conclusion}
The comparison between with and without graphene channel devices confirms that the crack issue in sol-gel Alumina layer post thermal annealing is due to underneath graphene layer. The possible mechanism for crack formation was thought to be restoring movement of graphene layer due to stress generated because of TEC mismatch amongst layers and shrinkage during annealing. Even-though DUV annealing with e-beam evaporated thin seed layer films (2.5 nm and 3.5 nm) were not able to resolve the crack issue. On other hand, 5 nm thick e-beam evaporated seed layer based devices did not show crack formation but, electrical testing confirms that the seed layer was not completely oxidised. Finally, we proposed novel two step seed layer deposition process (1.5 nm from e-beam and 1.5 nm from sputtering) with DUV annealing to cover side wall and restrict graphene movement with reduced shrinkage stress. The novel method not only resolved the crack issue but also retained Dirac point of the transistor. 
\section*{Acknowledgment}
This work was carried out under the Centre of Excellence in Nanoelectronics Phase-II project funded by the Ministry of Electronics and Information Technology, Government of India at the IIT Bombay Nanofabrication Facility. SEM imaging of seed layers was carried out at NCPRE and SEM images of GFETs are carried out at IRCC Central Facility at IIT Bombay

\ifCLASSOPTIONcaptionsoff
  \newpage
\fi

% trigger a \newpage just before the given reference
% number - used to balance the columns on the last page
% adjust value as needed - may need to be readjusted if
% the document is modified later
%\IEEEtriggeratref{8}
% The "triggered" command can be changed if desired:
%\IEEEtriggercmd{\enlargethispage{-5in}}

% references section

% can use a bibliography generated by BibTeX as a .bbl file
% BibTeX documentation can be easily obtained at:
% http://mirror.ctan.org/biblio/bibtex/contrib/doc/
% The IEEEtran BibTeX style support page is at:
% http://www.michaelshell.org/tex/ieeetran/bibtex/
%\bibliographystyle{IEEEtran}
% argument is your BibTeX string definitions and bibliography database(s)
%\bibliography{IEEEabrv,../bib/paper}
%
% <OR> manually copy in the resultant .bbl file
% set second argument of \begin to the number of references
% (used to reserve space for the reference number labels box)
\bibliographystyle{IEEEtran}
\bibliography{bare_jrnl.bib}

% biography section
% 
% If you have an EPS/PDF photo (graphicx package needed) extra braces are
% needed around the contents of the optional argument to biography to prevent
% the LaTeX parser from getting confused when it sees the complicated
% \includegraphics command within an optional argument. (You could create
% your own custom macro containing the \includegraphics command to make things
% simpler here.)
%\begin{IEEEbiography}[{\includegraphics[width=1in,height=1.25in,clip,keepaspectratio]{mshell}}]{Michael Shell}
% or if you just want to reserve a space for a photo:

%\begin{IEEEbiography}{Michael Shell}
%Biography text here.
%\end{IEEEbiography}

% if you will not have a photo at all:
%\begin{IEEEbiographynophoto}{John Doe}
%Biography text here.
%\end{IEEEbiographynophoto}

% insert where needed to balance the two columns on the last page with
% biographies
%\newpage

%\begin{IEEEbiographynophoto}{Jane Doe}
%Biography text here.
%\end{IEEEbiographynophoto}

% You can push biographies down or up by placing
% a \vfill before or after them. The appropriate
% use of \vfill depends on what kind of text is
% on the last page and whether or not the columns
% are being equalized.

%\vfill

% Can be used to pull up biographies so that the bottom of the last one
% is flush with the other column.
%\enlargethispage{-5in}

% that's all folks
\end{document}